\documentstyle[prl,aps]{revtex}

\input epsf
\begin{document}
\draft
\twocolumn[\hsize\textwidth\columnwidth\hsize\csname@twocolumnfalse\endcsname

\preprint{}
\title{Low Energy Quasiparticle Excitation in the Vortex State of
Borocarbide Superconductor YNi$_{2}$B$_{2}$C}
\author{K.~Izawa$^{1,2}$, A.~Shibata$^1$,
Yuji~Matsuda$^{1,2}$,Y.~Kato$^3$, H.~Takeya$^4$, K.~Hirata$^4$,
C.J.~van~der~Beek$^5$, and M.~Konczykowski$^5$}
\address{$^1$Institute for Solid State Physics, University of Tokyo,
Kashiwanoha 5-1-5, \\Kashiwa, Chiba 277-8581, Japan}
\address{$^2$CREST, Japan Science and Technology Corporation,
Kawaguchi, Saitama 332-0012, Japan}
\address{$^3$Department of Applied Physics, University of Tokyo, Tokyo
113-0033, Japan}
\address{$^4$National Research Institute for Metals, Tsukuba, Ibaraki
305-0047, Japan}
\address{$^5$Laboratoire des Solides Irradies, Ecole Polytechnique,
Palaiseau, 91128, France}

\maketitle

\begin{abstract}
We have measured the low temperature heat capacity $C_p$ and microwave
surface impedance $Z_s$ in the vortex state of YNi$_2$B$_2$C. In
contrast to conventional $s$-wave superconductors, $C_p$ shows a
nearly $\sqrt H$-dependence.  This $\sqrt H$-dependence persists even
after the introduction of the columnar defects which change the
electronic structure of the vortex core regime dramatically and
strongly disturb the regular vortex lattice.  On the other hand, flux
flow resistivity obtained from $Z_s$ is nearly proportional to $H$. 
Taken together, these results indicate that the vortex state of
YNi$_2$B$_2$C is fundamentally different from the conventional
$s$-wave counterparts, in that the delocalized quasiparticle states
around the vortex core play a much more important role, similar to
$d$-wave superconductors.
\end{abstract}
\pacs{74.70.Dd, 74.60.Ec, 74.25.Nf, 74.25.Jb}

]

\narrowtext

The borocarbide superconductors LnNi$_2$B$_2$C where Ln=(Y, Lu, Tm,
Er, Ho and Dy) exhibit a rich variety of interesting physical
properties.  In particular, the occurrence of superconductivity at
elevated temperatures \cite{cava}, the competition and coexistence of
the antiferromagnetic ordering and superconductivity \cite{canfield},
and the transition between a triangular and square vortex lattice
\cite{square} have attracted much attention.  In spite of extensive
studies on these subjects, however, one of the most fundamental
properties of the superconducting state, namely the quasiparticle (QP)
structure in the vortex state, is still controversial.  In fact,
recent measurements of low temperature heat capacity $C_p$ on
YNi$_2$B$_2$C and LuNi$_2$B$_2$C in the vortex state have shown that
$C_p$ clearly indicates the presence of the $\sqrt{H}$-term
\cite{nohara}.  In the conventional $s$-wave superconductors, $C_p$
should increase linearly with $H$ because all the QPs are trapped
within the vortex core with the radius of the coherence length $\xi$,
and hence the QP density of states (DOS) $N(H)$ is proportional to the
number of vortices; $N(H)\propto N_F\xi^2 H$ where $N_F$ is the DOS at
the Fermi level in the core \cite{caroli}.  Thus, the observed
$\sqrt{H}$-dependence of $C_p$ is strikingly in contrast to the
ordinary $s$-wave superconductors.  The $\sqrt H$-dependence of $C_p$
has been reported in some of the unconventional superconductors with
gap nodes in the QP energy spectrum such as high-$T_c$
YBa$_2$Cu$_3$O$_7$ \cite{moler} and heavy fermion UPt$_3$
\cite{ramirez} superconductors.  Nonlinear $H$-dependence close to the
$\sqrt H$ of $C_p$ is also observed in some $s$-wave clean
superconductors such as CeRu$_2$ \cite{hedo} and NbSe$_2$
\cite{nohara,sonier2}.

Although the unusual $H$-dependent $C_p$ of borocarbide
superconductors has been discussed in terms of several intriguing
models, the issue is still far from being settled.  For instance,
several authors proposed the shrinking of the vortex core with $H$
\cite{nohara,sonier2,sonier1,ichioka}, but the physical origin behind
this phenomenon is unclear.  Another group ascribed it to the
field-induced gap nodes \cite{hedo}, but this scenario is beyond the
applicability of the original argument\cite{brandt}.  Moreover, the
extended QP states outside the vortex core owing to a presupposed
$d$-wave symmetry have been invoked\cite{wang}.  There is, however, no
corroborative evidence for $d$-wave pairing.  Therefore, the
experimental clarification of the QP structure in the vortex state of
YNi$_2$B$_2$C is very much needed.  Further, it is crucial for
understanding the vortex lattice structure.  If a substantial portion
of the QPs extend in specific directions well outside the core, they
should play an important role in determining the superconducting
properties, including the vortex lattice structure, magnetization
$M(H)$, upper critical field $H_{c2}$, {\it etc}.  \cite{maki}. 
Nevertheless, these properties have been discussed in terms of an
ordinary $s$-wave superconductor with an anisotropic Fermi surface
with the use of the nonlocal London theory, {\it without taking into
account the effect of the extended QPs.} \cite{kogan}.

We stress that we can extract detailed information about the QP
spectrum only when we measure both the heat capacity and the surface
impedance $Z_{s}$ for the same sample.  In superconductors with gap
nodes, $C_p$ is essentially determined by the QPs excited {\it in the
node directions}\cite{volovik}.  Energy dissipation in the flux flow
state, on the other hand, occurs mainly in the ^^ ^^ normal regions"
created by the vortices.  Thus, the dissipative response is dominated
by Andr\'{e}ev bound states localized within vortex cores; those
states have momenta whose directions lie {\it away from the nodes}
\cite{kopnin}.  Therefore, the measurement of $Z_s$ is complementary
to the heat capacity.  In this Letter, we report measurements for both
$Z_s$ and $C_p$ in the vortex state of YNi$_2$B$_2$C. We have also
examined the effect of introducing columnar defects (CD) on $C_p$;
since the diameter of CD is comparable to $\xi \sim 70$~\AA~of
YNi$_2$B$_2$C and the inside of CD is semiconducting \cite{Nishida},
they should strongly influence the electronic structure of the
vortices.  As we shall discuss below, these results provide strong
evidence that the $\sqrt H$-term in $C_p$ originates mainly from the
Doppler shift of the delocalized QP spectrum due to the superfluid
electrons surrounding the vortex cores \cite{volovik}.  This suggests
that a strongly anisotropic $s$-wave state most likely applies to
YNi$_2$B$_2$C.

Single crystals of YNi$_{2}$B$_{2}$C with $T_{c} =$ 13.4~K were grown
by the floating zone method.  In all measurements, dc magnetic fields
$\mbox{\boldmath $H$}$ were applied parallel to the $c$ axis.  The
microwave surface impedance $Z_s = R_s +iX_s$, where $R_s$ and $X_s$
are the surface resistance and the reactance, respectively, were
measured by the cavity perturbation technique.  We used a cylindrical
Cu cavity with $Q \sim 3\times 10^{4}$ operated at 28.5~GHz in the
TE$_{011}$ mode.  The sample was placed at the center of the cavity
which is the antinode of the microwave magnetic field $\mbox{\boldmath
$H$}_{ac}$ that is parallel to the $c$-axis of the sample
($\mbox{\boldmath $H$}_{ac}\parallel \mbox{\boldmath $H$} \parallel
c$).  In this configuration, the vortex lines along the $c$ axis
respond to oscillatory driving currents within the $ab$ plane induced
by $\mbox{\boldmath $H$}_{ac}$.  We measured $C_{p}$ by the thermal
relaxation method.  Several single crystals nearly one millimeter in
size and $\sim 100 \mu$m in thickness along the $c$ axis were
irradiated at GANIL (Caen, France) with a 6.0-GeV Pb-ion beam aligned
parallel to the $c$-axis.  The samples were irradiated to the fluence
of $5\times10^{10}$ and $1\times10^{11}$ ions/cm$^2$, corresponding to
a dose-equivalent flux density of $B_{\Phi}$=1~T and 2~T,
respectively.  The resulting damage consisted of a random array of
amorphous columns, about 60-70\AA~ in diameter and continuous
throughout the thickness of the crystal, which we confirmed by the
transmission electron microscope image.  The CD act as strong pinning
centers of the vortices; at $B \alt B_{\Phi}$, a substantial portion
of the vortices are trapped inside the CD. Particularly below $\sim
0.2B_\phi$, almost all vortices are expected to be trapped
\cite{Indenbom}.  The CD do not, however, act as strong scattering
centers of the electrons, unlike magnetic impurities or point defects. 
In fact, after the irradiation, $M(H)$ showed a broad peak at $B \sim
B_{\Phi}$ due to the trapping of the vortices by the CD (inset of
Fig.1), but the increase in the resistivity was very small ( less than
10\%) and the $T_c$ did not change.

Figure 1 plots $C_p/T$ as a function of~$T^2$ for the pristine
YNi$_2$B$_2$C. We will first make a remark on the $T$-dependence of
$C_p$.  Obviously, the normal state in Fig.  1 ($C_p/T$ vs $T^2$) that
occurs as the superconductivity is suppressed by magnetic fields
deviates from a linear dependence.  Though this deviation may be
attributed to the low-energy optical phonons, it is not clear where
they originate from \cite{michor}.  Moreover, recent inelastic neutron
scattering experiments demonstrated an anomalous phonon branch which
strongly couples to the electronic excitation in the superconducting
state \cite{kawano}.  Therefore, the quantitative analysis of the
$T$-dependence of the electronic contribution obtained by subtracting
the phonon 
\begin{figure}
    \centerline{\epsfxsize 8cm \epsfbox{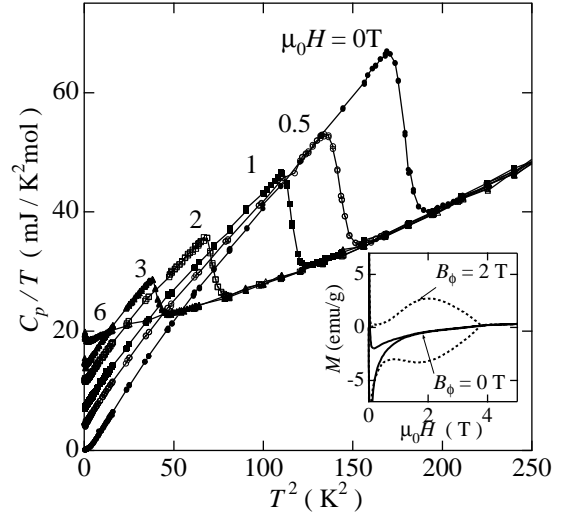}}
\caption{$C_p/T$ versus $T^2$ for the pristine YNi$_2$B$_2$C. All
measurements have been done in the field cooling conditions.  Inset:
$M-H$ curves for pristine and irradiated ($B_{\Phi}$=2~T) crystals at
5~K. While in the pristine crystal $M(H)$ is reversible in a wide
$H$-range, showing very weak pinning centers, $M(H)$ show a broad peak
at $B=B_{\Phi}$ in the irradiated crystal.  }
\end{figure}
\noindent terms from the total heat capacity is precarious.

Figure 2(a) depicts the $H$-dependence of the heat capacity for the
pristine YNi$_2$B$_2$C at a low temperature where the phonon
contribution is negligible.  In order to discuss the heat capacity
induced by a magnetic field, we plotted $\Delta C_p(H)/T =
(C_p(H)-C_p(0))/T$ in Fig.2(a).  A significant deviation from a linear
$H$-dependence is clearly observed.  The inset of Fig.  2(a) plots
$\Delta C_p(H)/T$ as a function of $\sqrt H$.  At a low field, $\Delta
C_p(H)/T$ increases with an upward curvature with respect to $\sqrt
H$.  However, at $\sqrt H \agt 0.25$~T$^{1/2}$ ($H \agt 70$~mT),
$\Delta C_p/T$ obviously increases linearly to $\sqrt H$.  This
$\sqrt H$-dependence persists close to $H_{c2}$.  Since the lower
critical field $H_{c1}$ determined by the $M$-$H$ curve is of the
order of 30~mT, the $\sqrt H$-dependent $\Delta C_p/T$ is observed in
almost the whole regime of the vortex state.  Figure 2(b) shows the
same plot for the irradiated YNi$_2$B$_2$C with $B_{\Phi}$=2~T. A
significant deviation from the linear $H$-dependence persists even
after the irradiation.  Surprisingly, $\Delta C_p/T$ is little
affected by the introduction of the CD (see the inset of Fig.2(b)). 
We obtained qualitatively similar results for YNi$_2$B$_2$C with
$B_{\Phi}$=1T. Before discussing these results, we will discuss $Z_s$
in the pristine YNi$_2$B$_2$C.

The inset of Fig.3 shows the $T$-dependence of $Z_s$ in the Meissner
phase.  Both $R_s$ and $X_s$ decrease rapidly with decreasing $T$
below $T_c$.  Figure 3 shows $Z_s$ as a function of $\sqrt{H}$ at
1.5~K. Let us quickly recall the behavior of $Z_s$ in type-II
superconductors.  In the Meissner phase, the microwave response is
purely reactive and $R_s \simeq 0$ and $X_s=\mu_0\omega\lambda_L$,
where $\mu_0$ is the vacuum permeability, $\omega$ is the microwave
frequency and $\lambda_L$ is the London penetra-
\begin{figure}
    \centerline{\epsfxsize 8cm \epsfbox{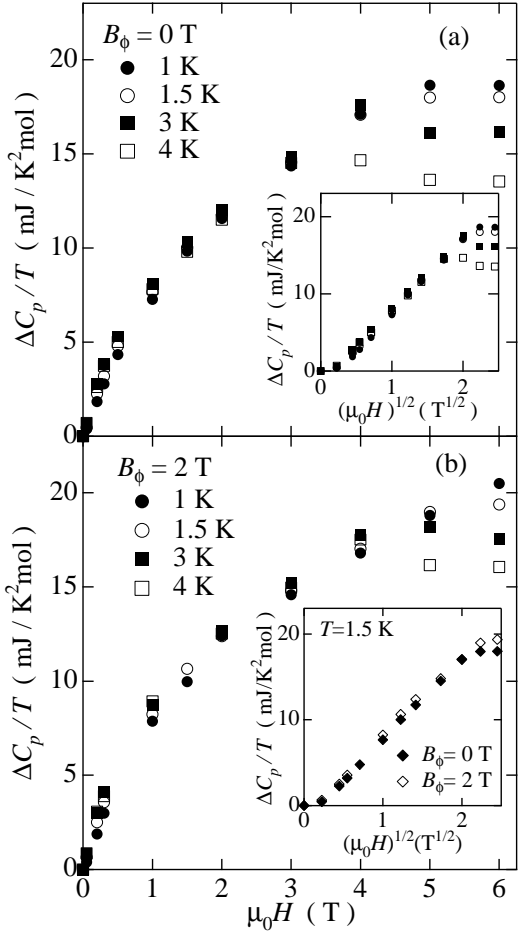}}
\caption{(a) $\Delta C_p(H)/T = 
\{C_{p}(H)-C_{p}(0)\}/T$ for the pristine YNi$_2$B$_2$C at low
temperatures.  Inset: $\Delta C_p/T$ as a function of $\sqrt H$.  (b)
The same plot for the irradiated crystal with 
$B_{\Phi}=2$~T\protect\cite{accuracy}. Inset:
The $H$-dependence of $\Delta C_p/T$ before and after the
irradiation.}
\end{figure}
\noindent tion length.  On the
other hand, the response is dissipative in the normal state and
$R_s=X_s=\mu_0\omega\delta/2$, where
$\delta=\sqrt{2\rho_n/\mu_0\omega}$ is the skin depth and $\rho_n$ is
the normal state resistivity.  We determined the absolute values of
$R_s$ and $X_s$ from the comparison with the dc resistivity and
assuming that $R_s=X_s$ in the normal state.  Using this procedure, we
obtained $\lambda_L\simeq 500$~\AA. In the vortex state, $Z_s$ is
determined by the vortex dynamics.  Since the present microwave
frequency is three orders of magnitude higher than the vortex pinning
frequency of YNi$_2$B$_2$C \cite{oxx}, a completely free flux flow
state is realized.  In the flux flow state, two characteristic length
scales, namely $\lambda_L$ and the flux flow skin depth $\delta_f\sim
\sqrt{2\rho_f/\mu_0\omega}$, appear in accordance with the microwave
field penetration.  At a low field, $\lambda_L$ greatly exceeds
$\delta_f$ ($\lambda_L \gg \delta_f$).  In this regime, $R_s$ and
$X_s$ are given as $R_s\sim\rho_f/\lambda_L$ and $X_s\sim
\mu_0\omega\lambda_L$.  On the other hand, at high fields where
$\delta_f$ greatly exceeds $\lambda_L$ ($\delta_f \gg \lambda_L$), the
viscous loss becomes dominant and the response is similar to the
normal state 
\begin{figure}
    \centerline{\epsfxsize 8cm \epsfbox{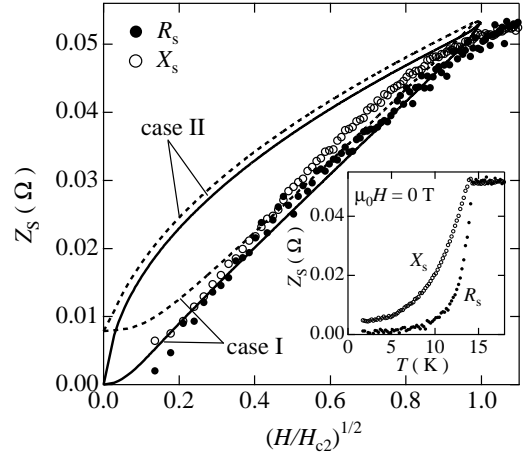}}
\caption{Inset: $T$-dependence of $Z_s=R_s+iX_s$ in the Meissner
phase.  Main panel: $R_s$ and $X_s$ as a function of
$(H/H_{c2})^{1/2}$ at 1.5~K. The solid and dashed lines represent the
result of the theoretical calculations of $R_s$ and $X_s$ by Eq.(1),
respectively, assuming two different $H$-dependences of $N(H)$; $N(H)
\propto H$ (case I) and $N(H) \propto \sqrt H$ (case II).  For the
detail, see text.}
\end{figure}
\noindent ($R_s\simeq X_s$) except that $\delta$ is replaced by
$\delta_f$.  In YNi$_2$B$_2$C this crossover occurs at a very low
field ($\mu_0 H\sim 20$~mT).  According to Coffey and Clem
\cite{coffey}, $Z_s$ in the flux flow state is expressed as
\begin{equation}
	Z_s=i\mu_{0}\omega\lambda_L\left[\frac{1-(i/2){\delta_f}^{2}/
\lambda_{L}^{2}}{1+2i\lambda_L^{2}/\delta_{qp}^2}\right]^{1/2},
\end{equation}
where $\delta_{qp}=\sqrt{2\rho_{qp}/\mu_0\omega}$ with the QP
resistivity $\rho_{qp}$ is the normal-fluid skin depth.  We analyzed
the data by means of Eq.(1).  We considered two different
$H$-dependences of $N(H)$, namely $N(H)\propto H$ (case I) and
$N(H)\propto \sqrt{H}$ (case II).  Case I corresponds to the
conventional Bardeen-Stephen relation, $\rho_f=(H/H_{c2})\rho_n$ and
$\lambda_L^2(H)=\lambda_L^2(0)/(1-H/H_{c2})$, while case II
corresponds to the vortex core shrinking, $\rho_f\sim
\sqrt{H/H_{c2}}\rho_n$ and
$\lambda_L^2(H)=\lambda_L^2(0)/(1-\sqrt{H/H_{c2}})$.  In both cases we
used $\lambda_L(0)=500$\AA \cite{lambda}.  The solid and dashed lines
represent the theoretical calculations of $R_s$ and $X_s$ by Eq.(1),
respectively.  Comparing the two cases, case I obviously describes the
data much better than case II. Small deviations of the fits from the
data in case I may be due to the fact that $\rho_f$ is not strictly
linear in $H$, as suggested by the time-dependent Ginzburg-Landau
theory \cite{larkin}.

The results of $Z_s$ and $\Delta C_p$ for the pristine and irradiated
YNi$_2$B$_2$C offer important clues for understanding the QP
structure.  Figure 3 shows that the linear $H$-dependent
Bardeen-Stephen relation yields a more consistent fit to the data than
another model does.  Since the QPs localized in the core mainly
contribute to the flux flow dissipation, this fact shows that the
number of the QPs trapped within each core is independent of $H$. 
Therefore, the scenario of the core shrinking with $H$ as an origin of
nonlinear $H$-dependent $C_p$ proposed by several groups
\cite{nohara,sonier2,sonier1,ichioka} is completely excluded.  In
addition and more importantly, the fact that the existence of the CD
with a comparable radius with $\xi$ little affects the $\Delta C_p$
implies that the QPs within the core radius $\xi$ are not important
for the total heat capacity in the pristine YNi$_{2}$B$_{2}$C. On the
basis of these results, we are led to conclude that in YNi$_2$B$_2$C
{\it the extended QP states around the vortex core play an important
role in determining the superconducting properties, similar to
$d$-wave superconductors}.

Moreover, the effect of the CD on the $\Delta C_p$ provides another
important piece of information about the microscopic origin of $\sqrt
H$-dependence.  It has been pointed out that in the presence of line
nodes there are two sources for the $\sqrt{H}$-dependent $N(H)$,
namely, the contributions from localized and delocalized fermions
\cite{volovik}, but which of the two dominates has thus far been left
unquestioned.  Here we shall attempt to address this issue.  One
arises from the localized QPs in the node directions.  Since the
extended QP wavefunction in the node directions is cut off by its
adjacent vortices, the area per a single vortex is proportional to the
intervortex distance $R \propto 1/\sqrt{H}$.  Then $N(H)$ is given as
$N(H) \propto N_F \xi R H \propto \sqrt{H}$.  The other arises from
the delocalized QPs.  In the presence of the supercurrent flow with
velocity \mbox{\boldmath $v$}$_{s}$, the energy spectrum of the
delocalized QPs is shifted by the Doppler effect as $E(\mbox{\boldmath
$p$})\rightarrow E(\mbox{\boldmath $p$})+\mbox{\boldmath $v$}_s \cdot
\mbox{\boldmath $p$}$.  In superconductors with line nodes such as
$d$-wave symmetry where DOS has a linear energy dependence in the bulk
($N(E)\propto E$), the Doppler-shifted QPs give rise to the finite DOS
at the Fermi level.  Then $N(H)$ is obtained by integrating the vortex
lattice cell $N(H) \propto R^{-2} \int_{\xi}^{R}$\mbox{\boldmath
$v$}$_s$(\mbox{\boldmath $r$}) $\cdot$ \mbox{\boldmath $p$}
$r~dr\propto \sqrt{H}$, where $\left|\mbox{\boldmath
$v$}_s(\mbox{\boldmath $r$})\right|=\hbar/2mr$ is the velocity of the
circulating supercurrents.  The very weak influence of the CD on
$\Delta C_p$ elucidates the fact that {\it the Doppler shift of the
delocalized QPs is mainly responsible for the $\sqrt H$ dependent
$C_p$}, because the contribution from the localized QPs in the node
directions should diminish when the vortex lattice structure is
destroyed.

At low fields ($H \alt 70$~mT $\approx H_{cr}$), the $\Delta C_p/T$
vs.  $\sqrt{H}$ curve shows an upward curvature.  There are two
origins for the deviation from the $\sqrt{H}$-behavior.  The first one
is the lower critical field $H_{c1}\sim$ 30~mT, which is in the same
order of $H_{cr}$.  The second one is the crossover from high field to
low field scaling.  Generally, the heat capacity of superconductors
with line nodes is proportional to $\sqrt{H}$ only at {\it high
fields} where the number of the Doppler-shifted QPs exceeds that of
thermally excited QPs.  The heat capacity at low fields should be
described by the zero-field scaling $C_p/(\gamma_n T)\sim k_B
T/\Delta$, where $\gamma_n$ and $\Delta$ are the Sommerfeld
coefficient in the normal state and superconducting energy gap,
respectively \cite{kopnin2}.  The crossover field from high field
scaling to low field scaling roughly estimated from the relation
$\sqrt{H/H_{c2}} \sim k_{\rm B}T/\Delta$ with $\mu_0 H_{c2}=5.5$~T,
$T_c=13.4$~K and $T=1.5$~K is $\sim$24~mT, which is also in the same
order of $H_{cr}$.  It should be noted that this behavior is also
reported in YBa$_2$Cu$_3$O$_{7-\delta}$ with $d$-wave symmetry
\cite{junod}.

The present experiments strongly suggest that the influence of the
extended QPs, which has been ignored by many authors, should be taken
into account when discussing the vortex lattice structure, $H_{c2}$
and $M(H)$.  In YNi$_2$B$_2$C, de Haas-van Alphen oscillations are
observed as low as $H_{c2}/5$ \cite{terashima}.  This unusual
phenomena may be related to the extended QPs.  It is tempting to
relate the observed relevance of the delocalized QPs to a
three-dimensional $d$-wave superconductivity \cite{wang}.  However,
recent doping studies on YNi$_2$B$_2$C show that the superconductivity
survives even in the dirty limit \cite{nohara,choen}.  This robustness
of the superconductivity against impurities makes a $d$-wave state
unlikely \cite{borkowski}.  Therefore, we believe that a strongly
anisotropic $s$-wave state in which $N(E) \propto E$ is similar to a
$d$-wave state is most likely for YNi$_2$B$_2$C.

We finally comment on the tunneling measurement by Ekino {\it et al.}
which reports a full BCS gap in YNi$_2$B$_2$C \cite{ekino1}.  We
believe that their break junction method selects a specific
$q$-direction in the gap function; in fact the same group reported a
full BCS gap even in $d$-wave high-$T_c$ cuprates \cite{ekino2}.

In summary, we have studied the heat capacity and surface impedance of
pristine and irradiated YNi$_2$B$_2$C at low temperatures.  We
provided strong evidence that the Doppler shift of the delocalized QP
spectrum around the vortex core plays an important role in determining
the superconducting properties of YNi$_{2}$B$_{2}$C.

We thank N.~Chikumoto, J.~Clem, T.~Hanaguri, R.P.~Huebener,
K.~Maki, H.~Takagi, A.~Tanaka and H.~Yoshizawa for their helpful
discussions.


\begin{thebibliography}{99}
\bibitem{lambda}Our $\lambda_L(0)$ is nearly half of $\lambda_{L}$
reported by $\mu$SR measurement \cite{msr}.  However the
$H$-dependence of $Z_s$ is little affected by the absolute value of
$\lambda_{L}$ except for the low field regime.
\bibitem{accuracy} Since the mass of the irradiated crystal is very
small, the accuracy of $C_{p}$ in this work is about 5~\%.  The
expected change of the normal-state $C_{p}$ due to the irradiation is
smaller than the accuracy.
\end{thebibliography}
\end{document}